\documentclass[lettersize,journal]{IEEEtran}
\usepackage{amsmath,amsfonts}
\usepackage{algorithmic}
\usepackage{algorithm}
\usepackage[caption=false,font=normalsize,labelfont=sf,textfont=sf]{subfig}
\usepackage{textcomp}
\usepackage{stfloats}
\usepackage{url}
\usepackage{verbatim}
\usepackage{graphicx}
\usepackage{cite}
\usepackage{booktabs}
\usepackage{tabularx}
\usepackage{array}
\usepackage{makecell}
\usepackage{threeparttable}
\begin{document}

\title{Measurement-Based Bitrate--Energy--Quality Analysis of Neural Audio Codec Decoders on Laptop and Phone Platforms}

\author{Seunghyeon Shin, Seokjin Lee~\IEEEmembership{(Member, IEEE)}
\thanks{This paper was produced by the IEEE Publication Technology Group. They are in Piscataway, NJ.}
\thanks{Manuscript received April 19, 2021; revised August 16, 2021.}}

\markboth{Journal of \LaTeX\ Class Files,~Vol.~14, No.~8, August~2021}%
{Shell \MakeLowercase{\textit{et al.}}: A Sample Article Using IEEEtran.cls for IEEE Journals}


\maketitle

\begin{abstract}

Neural audio codecs can achieve similar objective quality at lower bitrates than conventional codecs, but their decoder-side computational cost may offset this bitrate advantage on battery-powered client devices. This paper presents a measurement-based rate–energy–quality analysis of four neural audio codecs–EnCodec, DAC, HILCodec, and SNAC–and two conventional baselines, AAC-LC and Opus, on laptop and phone platforms. Speech and music are evaluated separately using the original-reference ViSQOL protocol. For the main comparison, operating points are matched by selecting the measured point nearest to the midpoint of the common overlap in treatment-level mean ViSQOL. A secondary analysis includes only explicitly evaluated bitrate settings. Decoder energy is measured as idle-subtracted energy per second of audio (J/s), and results are summarized as the median of three repeated runs for execution-valid runtime and device paths. Pairwise break-even transmission-energy thresholds are then derived analytically from the measured decoder energy and bitrate. At the matched-quality operating points, the evaluated neural codecs achieved similar ViSQOL scores at lower bitrates but generally required more decoder-side energy than the applicable conventional codecs. EnCodec produced the lowest neural break-even thresholds in both matched-quality cohorts on both platforms. By contrast, several DAC and SNAC comparisons on the Phone XNNPACK CPU path exceeded 200 mJ/kbit, and their full-band Phone configurations required more than 1 s to decode 1 s of audio. These results show that lower bitrate alone does not guarantee an energy benefit: deployment efficiency also depends on decoder complexity, the effective runtime and device mapping, execution validity, accelerator availability, and the transmission-energy coefficient.
\end{abstract}

\begin{IEEEkeywords}
Audio coding, neural audio codec, decoder energy, on-device inference, rate--energy--quality analysis, battery-powered devices.
\end{IEEEkeywords}

\section{Introduction}

\IEEEPARstart{N}{eural} audio codecs learn an analysis-quantization-synthesis pipeline from waveform data. An encoder maps the waveform to a lower-rate latent sequence, one or more vector quantizers represent this sequence using discrete indices, and a neural decoder reconstructs the waveform from the transmitted indices. Recent progress in residual and multi-scale quantization, perceptually oriented training, and streamable model design has enabled competitive speech and music quality at low bitrates~\cite{zeghidour2021soundstream,defossez2022highfi, kumar2023high,ahn2024hilcodec,siuzdak2024snac}. These developments have increased interest in neural codecs for bandwidth-constrained media delivery, but they also have made decoder complexity an important consideration for client-side deployment.

A recent survey describes the progression from conventional and hybrid speech coding to end-to-end neural audio codecs~\cite{kim2025neural}. Comparative frameworks such as Codec-SUPERB evaluate codec outputs using signal-level metrics and downstream tasks~\cite{wu2024codec}. Most neural codec studies focus on bitrate, objective or subjective quality, model complexity, and decoding latency. Although these indicators are useful for deployment analysis, they do not directly quantify decoder-side energy consumption or determine whether an accelerator-targeted graph is actually executed on the requested device. 

This distinction is particularly relevant to repeated consumer playback. In on-demand streaming, each content item is typically encoded once at the server, whereas transmission and client-side decoding occur during every playback session. For downloaded media, transmission occurs when the content is delivered, while decoding is performed during playback. A lower bitrate can therefore reduce the modeled transmission-energy term, while the decoder can increase client-side energy consumption. Whether neural coding provides an overall energy benefit thus depends on whether the transmission-energy savings associated with the lower bitrate compensate for the additional decoder-side energy on the effective execution path.

Earlier work on embedded systems measured the energy consumption of conventional audio decoding across codecs and operating settings~\cite{lin2008energy}. In video coding, decoder-energy models derived from bitstream features have been incorporated into encoder-side rate-distortion optimization~\cite{herglotz2016modeling,herglotz2017decoding}. These studies provide a system-level basis for considering decoder energy together with coding performance. From this perspective, the present work provides a measurement-based rate–energy–quality evaluation of neural and conventional audio decoder paths that combines original-reference objective-quality measurements with evidence of effective device execution.

This work makes three contributions. First, it defines a decoder-only measurement boundary, platform-specific telemetry, and idle-subtracted energy normalization for neural and conventional audio codecs on Laptop and Phone. Second, it defines the main matched-quality comparison by selecting the measured operating point nearest to the midpoint of the common ViSQOL overlap and complements this comparison with an exact bitrate-specified analysis that neither interpolates nor substitutes unavailable operating points. Third, it uses evidence of effective target execution to screen runtime paths and derives analytic pairwise break-even thresholds that quantify when bitrate savings offset additional decoder-side energy.

\section{Related Work}

\subsection{Audio Codec Architectures and Operating Points}

\begin{figure*}[t]
    \centering
    \includegraphics[width=0.98\textwidth]{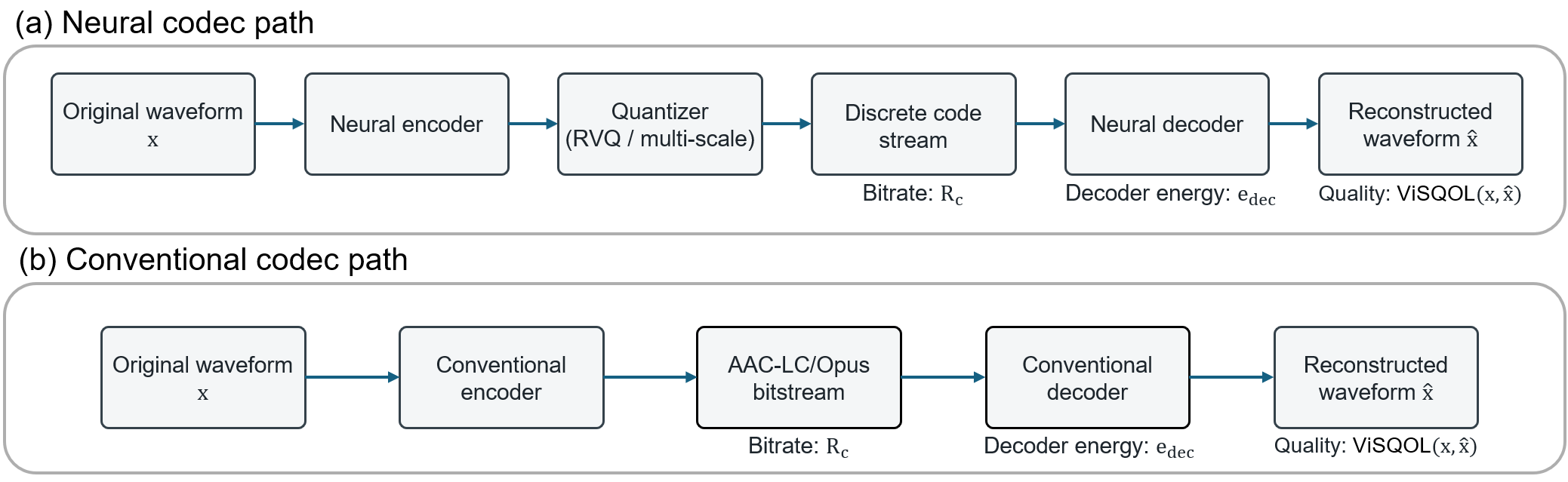}
    \caption{Neural and conventional codec paths. \(R_c\), \(e_{\mathrm{dec}}\), and ViSQOL denote transmitted bitrate, decoder-side energy, and original-reference quality, respectively; encoding and quantization are outside the energy measurement boundary.}
    \label{fig:codec_paths}
\end{figure*}

Most end-to-end neural audio codecs use a waveform encoder, a discrete quantizer, and a learned waveform decoder, as summarized in Fig.~\ref{fig:codec_paths}. SoundStream uses a fully convolutional encoder-decoder, residual vector quantization (RVQ), and quantizer dropout to support variable-rate operation with a single model~\cite{zeghidour2021soundstream}. EnCodec follows a streamable RVQ-based design and uses a multiscale spectrogram adversary and loss balancing~\cite{defossez2022highfi}. DAC extends RVQGAN-based compression through improved codebook usage, periodic inductive bias, and revised reconstruction and adversarial objectives~\cite{kumar2023high}. HILCodec uses causal and depthwise-separable convolutions to support streaming and low-complexity decoding~\cite{ahn2024hilcodec}, whereas SNAC distributes quantization across multiple temporal resolutions~\cite{siuzdak2024snac}.

For EnCodec, DAC, and HILCodec, the evaluated rate ladders are obtained mainly by varying the active RVQ depth. SNAC instead distributes the retained codes across multiple temporal resolutions. However, reducing the transmitted code set does not necessarily reduce the main waveform-synthesis operations performed by the decoder. Decoder cost may therefore change much less than bitrate or the number of active codebooks. For this reason, decoder cost is measured directly rather than inferred from the quantizer configuration.

AAC-LC and Opus serve as conventional baselines for comparison. AAC-LC uses an MDCT-based transform-coding framework with psychoacoustic bit allocation, quantization, and entropy coding~\cite{bosi1997iso}. Opus combines the linear-prediction tools of SILK with the MDCT-based CELT layer to support speech and music coding~\cite{valin2012opus}. The applicable operating bitrates of both codecs depend on the application and codec configuration. In this study, AAC-LC and Opus were evaluated using nominal encoder settings spanning 12–128 and 12–96~kbps, respectively, as listed in Table~\ref{tab:codec_instances}. Their mature decoder implementations provide practical reference points for client-side energy measurements.

\subsection{Decoder Energy and Runtime Mapping}

Prior neural codec studies primarily report rate–quality performance, listening-test results, parameter counts, operation counts, or decoding latency. AudioDec reports CPU and GPU latency~\cite{wu2023audiodec}, and HILCodec reports complexity and real-time-feasibility results~\cite{ahn2024hilcodec}. These indicators characterize computational demand and execution time, whereas decoder-side energy requires measurement on a specified client platform and execution path. Earlier embedded-system work measured energy consumption for conventional audio decoding across codec, bitrate, and sampling settings~\cite{lin2008energy}. The present study extends this measurement perspective to neural and conventional decoders across multiple runtime and device paths and also records effective execution.

Video coding has developed explicit methods for accounting for decoder energy. Prior work estimated decoder energy from features of standardized video bitstreams and incorporated the estimates into encoder-side rate--distortion optimization~\cite{herglotz2016modeling,herglotz2017decoding}. Mobile-inference studies further show that observed behavior depends on the combination of model, runtime, and accelerator ~\cite{janapa2022mlperf,tan2023deep}. Building on these system-level perspectives, the present work directly measures decoder-side energy for neural and conventional audio codec paths, selects discrete operating points in original-reference ViSQOL space, records evidence of effective execution, and retains transmission energy as a parametric sensitivity term.

\section{Analysis Framework} \label{sec3}

The analysis considers one-to-many playback. Server-side encoding is performed once per content item and is excluded from the model. The analysis combines measured client-side decoder energy with modeled transmission energy for the applicable delivery and playback events.

Because equal bitrate does not imply equal perceptual quality, the main analysis uses overlap-based matched-quality selection in ViSQOL space. A secondary bitrate-specified analysis includes only explicitly evaluated operating points, without interpolation or nearest-bitrate substitution.

For a codec instance \(c\), execution configuration \(p\), transmission-energy regime \(N\), and playback duration \(T\), the total energy is modeled as follows:

\begin{equation}
E_{\mathrm{total}}(c,p,N,T)
=
E_{\mathrm{dec}}(c,p,T)
+
E_{\mathrm{tx}}(R_c,N,T),
\end{equation}

where \(E_{\mathrm{dec}}\) is the client-side decoder energy, \(E_{\mathrm{tx}}\) is the transmission energy, and \(R_c\) is the bitrate of codec instance \(c\). The execution configuration \(p\) includes the platform, runtime, target device, and precision where applicable.

The benchmark records platform-specific telemetry samples and timestamps during the idle-baseline and measured-decoding intervals, together with the run duration \(\Delta t_{\mathrm{run}}(c,p)\) and decoded-audio duration \(T_{\text{audio}}(c,p)\). The run energy \(E_{\mathrm{run}}(c,p)\) and idle power \(P_{\mathrm{idle}}\) are derived from the applicable telemetry as detailed in Section~\ref{sec5}. The idle-subtracted decoder energy is then defined as:

\begin{equation}
E_{\mathrm{dec}}(c,p)
=
E_{\mathrm{run}}(c,p)
-
P_{\mathrm{idle}}\Delta t_{\mathrm{run}}(c,p).
\end{equation}

In order to compare runs with different decoded audio durations, we normalize this value using the decoded audio duration:

\begin{equation}
e_{\mathrm{dec}}(c,p)
=
\frac{E_{\mathrm{dec}}(c,p)}
{T_{\mathrm{audio}}(c,p)}.
\end{equation}

Here, \(e_{\mathrm{dec}}\) denotes idle-subtracted decoder-side energy per unit decoded-audio duration. Because it is expressed in joules per second of audio, it is numerically equivalent to the average additional decoding power, in watts, under real-time (1$\times$) playback. This interpretation represents a normalization of measured energy and does not imply that every evaluated path meets real-time latency requirements. For a playback duration \(T\), we assume steady-state decoding and approximate

\begin{equation}
E_{\mathrm{dec}}(c,p,T)
\approx
e_{\mathrm{dec}}(c,p)T.
\end{equation}

Transmission energy is modeled using a transmission-energy coefficient \(\alpha_N\). When \(R_c\) is expressed in bits per second, \(\alpha_N\) has units of joules per bit. The transmission energy is approximated as

\begin{equation}
E_{\mathrm{tx}}(R_c,N,T)
=
\alpha_N R_c T.
\end{equation}

The total energy can therefore be written as

\begin{equation}
E_{\text{total}}(c,p,N,T)
=
T\left(e_{\mathrm{dec}}(c,p)+\alpha_N R_c\right).
\end{equation}

For the reported pairwise comparisons, \(A=(c_A,p_A)\) denotes the lower-bitrate neural configuration, and \(B=(c_B,p_B)\) denotes the corresponding conventional reference. We define \(\Delta e_{A,B} = e_{\mathrm{dec}}(c_A,p_A)-e_{\mathrm{dec}}(c_B,p_B)\) as the decoder-energy difference and \(\Delta R_{A,B}=R_{c_B}-R_{c_A}\) as the bitrate reduction. Equating the modeled total energies of configurations \(A\) and \(B\) yields

\begin{equation}
\alpha^*_{A,B}
=
\frac{\Delta e_{A,B}}{\Delta R_{A,B}}
=
\frac{
e_{\mathrm{dec}}(c_A,p_A)-e_{\mathrm{dec}}(c_B,p_B)
}
{
R_{c_B}-R_{c_A}
}.
\label{eq:break_even}
\end{equation}

Here, \(\alpha^*_{A,B}\) is the transmission-energy coefficient at which the modeled transmission-energy saving equals the decoder-energy difference. Configuration \(A\) has lower modeled transmission-plus-decoding energy when \(\alpha_N>\alpha^*_{A,B}\).

Published measurements of LTE, 5G, and Wi-Fi systems show that transmission energy depends strongly on access technology, throughput, link conditions, and radio state~\cite{huang2012close,xu2020understanding,sun2014modeling}. Accordingly, \(\alpha_N\) is treated as a parametric transmission-energy coefficient rather than as a measured constant for a particular wireless interface.  With \(e_{\mathrm{dec}}\) expressed in J/s of audio and bitrate in kbit/s, \(\alpha_N\) and \(\alpha^*_{A,B}\) are reported in mJ/kbit. Thresholds above 200~mJ/kbit are annotated with their exact values at the upper plotting limit.

Latency and fallback ratio are reported only as execution diagnostics. Latency is defined as the wall-clock time of one measured decoding iteration after warm-up and is not used as a primary metric. The fallback ratio is the share of profiled execution time assigned to non-target devices in accelerator-targeted runs. Paths that fail or do not satisfy the execution-validity criteria are excluded from the main comparison and reported separately as diagnostic results.

\section{Experimental Setup}\label{sec4}

\subsection{Hardware Platforms}\label{sec4_1}

We evaluated decoder energy on two client platforms, denoted Laptop and Phone. Table~\ref{tab:hardware_platforms} summarizes their hardware characteristics and measurement conditions.

The Laptop platform was a laptop-class system with CPU, GPU, and NPU resources integrated within a single processor package. Energy measurements were obtained from processor package-level telemetry. All Laptop measurements were performed with external power connected to maintain a stable platform state.

The Phone platform was a smartphone-class system based on a mobile SoC with integrated CPU, GPU, and NPU resources. For this platform, energy was obtained from battery-discharge telemetry exposed through the BatteryManager API~\cite{android_batterymanager}. During Phone measurements, the device operated on battery power with airplane mode enabled. Bluetooth, location services, and the always-on display were disabled, and the screen was kept on at a fixed minimum brightness.

The two platforms expose different telemetry layers. Laptop measurements reflect processor package-level energy, whereas Phone measurements reflect device-level battery-discharge energy. Absolute energy values are therefore analyzed only within each platform.

\begin{table}[t]
\centering
\caption{Platform characteristics and measurement conditions used for decoder benchmarking.}
\label{tab:hardware_platforms}
\small
\renewcommand{\arraystretch}{1.15}
\begin{threeparttable}
\begin{tabularx}{\linewidth}{@{}lXX@{}}
\toprule
\textbf{Item} & \textbf{Laptop} & \textbf{Phone} \\
\midrule

Device class
& Laptop-class client system
& Smartphone-class client system \\

Compute platform
& Integrated processor with CPU, GPU, and NPU resources
& Mobile SoC with CPU, GPU, and NPU resources \\

Memory
& 32 GB LPDDR5X
& 12 GB LPDDR5X \\

Operating system
& Windows 11 25H2
& Android 16 \\

Energy telemetry
& Processor package-level telemetry
& BatteryManager discharge telemetry \\

Power state
& Plugged in
& Unplugged \\

\bottomrule
\end{tabularx}
\end{threeparttable}
\end{table}

\subsection{Software Stack and Runtime Configuration}\label{sec4_2}

All neural audio codec models were exported to ONNX. FP32 and FP16 model artifacts were generated separately. ONNX export and laptop-side benchmarking were performed on the same laptop hardware, but in separate Python virtual environments.

On Laptop, Python 3.11 was used. Neural codec decoders were executed with ONNX Runtime 1.24.2~\cite{onnxruntime} and OpenVINO 2025.4.1~\cite{openvino}. Conventional codec decoders were executed with FFmpeg 8.0.1~\cite{ffmpeg}. The ONNX Runtime CPU path used the CPUExecutionProvider, and the GPU-targeted ONNX Runtime path used the DirectML ExecutionProvider. OpenVINO CPU, GPU, and NPU plugin paths were evaluated independently.

On Phone, the ONNX models were executed in a native mobile benchmarking application using ONNX Runtime 1.24.2~\cite{onnxruntime}. The Phone CPU path used the XNNPACK ExecutionProvider~\cite{xnnpack}. The accelerator-targeted path used the QNN ExecutionProvider, and the requested backend type was recorded as part of the execution-path metadata. Conventional codec decoders on Phone were executed through the MediaCodec system decoder path.

Only execution-valid paths were included in the main comparison. A path was considered execution-valid when session creation succeeded and the measured decoding completed. Accelerator-targeted paths were excluded when session creation or decoding failed, or when runtime evidence showed that the requested target was not effectively used. Partial non-target execution was retained and classified as mixed execution.
Table~\ref{tab:runtime_paths} summarizes the runtime paths considered in this study.

\begin{table}[t]
\centering
\caption{Runtime paths considered in the decoder benchmark.}
\label{tab:runtime_paths}
\small
\setlength{\tabcolsep}{4pt}
\renewcommand{\arraystretch}{1.12}
\begin{threeparttable}
\begin{tabularx}{\linewidth}{@{}
>{\raggedright\arraybackslash}p{0.16\linewidth}
>{\raggedright\arraybackslash}p{0.18\linewidth}
>{\raggedright\arraybackslash}X
>{\raggedright\arraybackslash}p{0.18\linewidth}
@{}}
\toprule
\textbf{Platform} &
\textbf{Codec type} &
\textbf{Runtime path} &
\textbf{Main comparison} \\
\midrule

Laptop
& Neural
& ONNX Runtime CPUExecutionProvider
& Included \\
Laptop
& Neural
& ONNX Runtime DirectML EP, GPU-targeted
& Excluded \\

Laptop
& Neural
& OpenVINO CPU plugin
& Included \\

Laptop
& Neural
& OpenVINO GPU plugin
& Included; mixed execution reported when observed \\

Laptop
& Neural
& OpenVINO NPU plugin
& Included \\

Laptop
& Conventional
& FFmpeg native decoder path
& Included \\

Phone
& Neural
& ONNX Runtime XNNPACK EP, CPU path
& Included \\

Phone
& Neural
& ONNX Runtime QNN EP, accelerator-targeted path
& Excluded \\

Phone
& Conventional
& MediaCodec system decoder path
& Included \\

\bottomrule
\end{tabularx}
\end{threeparttable}
\end{table}

\subsection{Codec Instances and Evaluated Operating Points}
\label{sec4_3}

We evaluated four neural audio codec families (EnCodec, DAC, HILCodec, and SNAC) and two conventional codec baselines, AAC-LC and Opus. For EnCodec, DAC, and HILCodec, the evaluated operating points were generated by varying the active RVQ depth within each released model instance. For SNAC, additional low-rate operating points were obtained by retaining only a subset of the scales or codebooks in the released checkpoints. These configurations are therefore treated as retained-scale-derived operating points rather than as separately trained and officially released low-bitrate checkpoints.

AAC-LC and Opus operating points were defined by encoder bitrate presets. Table~\ref{tab:codec_instances} summarizes the codec instances and bitrate settings evaluated in this study. Decoder inputs followed each codec’s native hop or frame settings and were held constant across runtime paths.

\begin{table}[t]
\centering
\caption{Codec instances and evaluated bitrate settings.}
\label{tab:codec_instances}
\scriptsize
\setlength{\tabcolsep}{3pt}
\renewcommand{\arraystretch}{1.05}
\begin{tabularx}{\columnwidth}{@{}lX@{}}
\toprule
\textbf{Codec} &
\textbf{Instances and evaluated bitrate settings (kbps)} \\
\midrule
EnCodec
& 24 kHz: 1.5, 3, 6, 12, 24; 48 kHz: 3, 6, 12, 24 \\

DAC
& 16 kHz: 1, 2, 4, 6; 24 kHz: 1.5, 3, 6, 12, 18, 24; 44.1 kHz: 1.72, 3.45, 5.17, 7.75 \\

HILCodec
& Speech 24 kHz: 1.5, 3, 6; music 24 kHz: 1.5, 3, 6, 9 \\

SNAC
& 24 kHz speech: 0.141, 0.422, 0.984; 32 kHz: 0.125, 0.375, 0.875, 1.875; 44.1 kHz: 0.172, 0.517, 1.206, 2.584 \\

AAC-LC
& 16, 24, 32, 44.1, 48 kHz: 12, 16, 24, 32, 48, 64, 96, 128 \\

Opus
& 16, 24, 32, 44.1, 48 kHz: 12, 16, 24, 32, 48, 64, 96 \\
\bottomrule
\end{tabularx}
\end{table}

\subsection{Evaluation Datasets, Quality Protocol, and Operating-Point Selection}

Quality evaluation was performed using a fixed set of evaluation clips for each domain. The speech domain used VCTK~\cite{yamagishi2019cstr} and was evaluated using the ViSQOL speech mode~\cite{chinen2020visqol}. The music domain used MUSDB18-HQ~\cite{MUSDB18HQ} and was evaluated using the ViSQOL audio mode~\cite{hines2015visqolaudio}. Speech and music were treated as separate perceptual domains, and ViSQOL scores were not pooled across them.

For each domain, we constructed 100 evaluation clips, each 10 s in duration. For VCTK, utterances shorter than 10 s were concatenated with other utterances from the same speaker when necessary. For MUSDB18-HQ, 10 s excerpts were selected while avoiding silent regions. Each evaluation clip was first resampled to the native sampling rate of the target codec instance before codec processing.

For ViSQOL evaluation, both the reference and decoded signals were resampled to the canonical sampling rate of the selected ViSQOL mode. Speech was evaluated at 16 kHz, whereas music was evaluated at 48 kHz. The main quality protocol used the original-reference setting. The reference signal was the original evaluation clip resampled to the canonical rate. The decoded signal was reconstructed at the codec’s native sampling rate and then resampled to the same canonical rate before evaluation. This procedure incorporates bandwidth loss into the measured quality degradation.

Clip-level ViSQOL scores were aggregated by domain, codec instance, and operating point to construct domain-specific rate--quality curves. HILCodec speech and music configurations were evaluated in their corresponding domains. The SNAC 24-kHz speech-targeted checkpoint was evaluated only in the speech domain.

We used two operating-point selection protocols. The main protocol was overlap-based matched-quality selection. A comparison cohort was defined as a set of codec instances that share the same domain and a comparable bandwidth class. For each cohort, we computed the common overlap interval of the treatment-level mean ViSQOL values. The midpoint of this interval was used as the quality anchor. For each codec instance, the matched operating point was selected as the nearest discrete point to this anchor.

As a secondary protocol, we used bitrate-specified energy evaluation. Only explicitly evaluated operating points were included; no interpolation, neighboring-bitrate substitution, or nearest-bitrate matching was used. A codec without an evaluated operating point was omitted from that condition. Panels were shown only when at least two codec series remained after this filtering. Under this display rule, the full-band music figure reports 12 and 24~kbps; the 3 and 6~kbps conditions were omitted because only EnCodec was available.

\section{Measurement Protocol}
\label{sec5}

\subsection{Benchmark Procedure}
\label{sec5_1}



\begin{figure}[t]
    \centering
    \includegraphics[width=\linewidth]{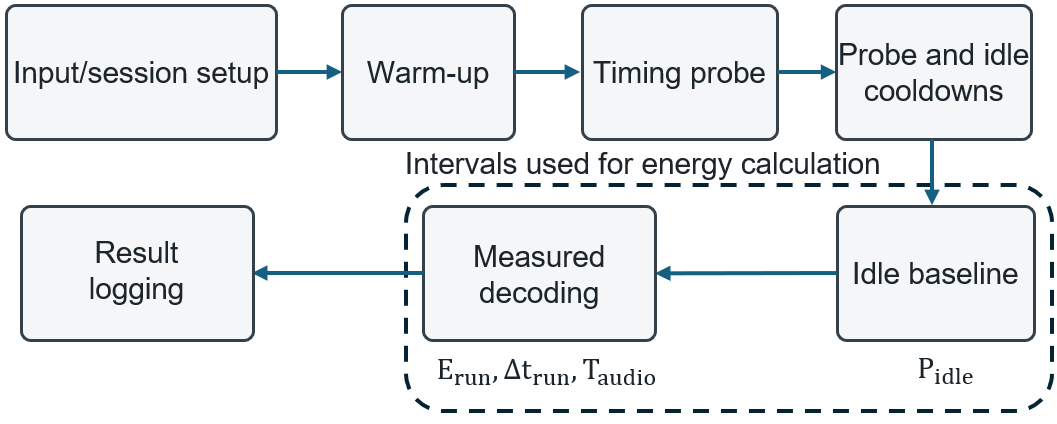}
    \caption{Benchmark sequence and energy-calculation intervals. The idle baseline and measured decoding provide the quantities used for idle subtraction.}
    \label{fig:benchmark_sequence}
\end{figure}

Fig.~\ref{fig:benchmark_sequence} summarizes the benchmark sequence and energy-measurement intervals. Benchmark inputs comprised 32 pre-generated 1-s artifacts supplied in a fixed cyclic order. Neural paths received code indices and optional scale values, whereas conventional paths received preloaded compressed inputs. The measured energy window covered steady-state decoder execution through reconstructed audio output and excluded setup, diagnostic profiling, and post-run result logging.

Each configuration underwent session initialization, warm-up, execution-time probing, probe cooldown, idle cooldown, idle-baseline acquisition, and measured decoding. The probe estimate was used to increase the iteration count when necessary to satisfy the platform-specific minimum measurement duration. Provider coverage and fallback profiling were collected, when supported, in a separate diagnostic pass. Session-creation and execution failures were retained as failure records.

Table~\ref{tab:benchmark_params} summarizes the benchmark parameters used for Laptop and Phone. Both platforms followed the same benchmark workflow, with timing and iteration parameters selected separately based on their sustained performance and thermal characteristics.

\begin{table}[t]
\centering
\caption{Benchmark parameters used for measured decoder runs.}
\label{tab:benchmark_params}
\scriptsize
\setlength{\tabcolsep}{3pt}
\renewcommand{\arraystretch}{1.05}
\begin{tabular}{@{}p{0.47\columnwidth}cc@{}}
\toprule
\textbf{Parameter} & \textbf{Laptop} & \textbf{Phone} \\
\midrule
Warm-up iterations & 300 & 30 \\
Probe iterations & 100 & 150 \\
Probe cooldown & 20 s & 30 s \\
Idle cooldown & 180 s & 240 s \\
Idle-baseline duration & 180 s & 240 s \\
Base measured iterations & 3000 & 300 \\
Minimum measured window & 120 s & 180 s \\
\bottomrule
\end{tabular}
\end{table}

The measured iteration count was increased automatically when the base iteration count did not satisfy the minimum measured window estimated from the probe run.

\subsection{Energy Measurement}
\label{sec5_2}


The common recorded quantities were telemetry samples and timestamps, run duration, decoded-audio duration, and execution status. Platform-specific telemetry was used to obtain \(E_{\mathrm{run}}\) and \(P_{\mathrm{idle}}\), from which \(E_{\mathrm{dec}}\) and \(e_{\mathrm{dec}}\) were derived as defined in Section~\ref{sec3}. Unless otherwise stated, the main comparison used \(e_{\mathrm{dec}}\), the idle-subtracted decoder-side energy normalized by decoded-audio duration.

On Laptop, processor package-level power telemetry was obtained through LibreHardwareMonitor~\cite{librehardwaremonitor}. Power samples were treated as interval-average values between consecutive sensor updates, and energy was computed by integrating these samples over the measured interval. The effective measurement boundaries were determined using the midpoint timestamps surrounding the start and end updates. Idle-baseline power was estimated from repeated 5-s idle sub-intervals. The main Laptop energy metric was processor package-level net energy after idle-baseline subtraction.

On Phone, current and voltage telemetry exposed through the Android BatteryManager API~\cite{android_batterymanager} were sampled at 1-s intervals. Discharge power was computed from the current--voltage samples and integrated using the trapezoidal rule. All Phone energy results reported in this study were obtained from these samples. Idle-baseline power was estimated from repeated 5-s idle sub-intervals and subtracted from the measured run energy.

Absolute energy values were interpreted only within each platform because Laptop and Phone use different telemetry layers.

\subsection{Result Aggregation}
\label{sec5_3}

For each execution configuration, three repeated benchmark runs were performed. The representative decoder-energy value was the median of the three runs, and the observed minimum and maximum were retained as run-to-run diagnostics. Diagnostic profiling passes were conducted separately and were not treated as independent energy measurements. Latency was summarized in the same manner and used only as an execution-time diagnostic.

Quality-based operating-point selection used treatment-level mean ViSQOL values. The 95\% bootstrap confidence intervals were obtained from 10,000 source-track-level bootstrap resamples and were used only to visualize uncertainty in the rate--quality curves; they were not used for operating-point selection~\cite{hesterberg2011bootstrap}. For each matched-quality cohort, the midpoint of the common mean-quality overlap was used as the anchor, and the nearest measured operating point was selected separately for each codec instance.

The break-even analysis used the selected bitrate and platform-specific median \(e_{\mathrm{dec}}\). Pairwise thresholds were calculated analytically from ~\eqref{eq:break_even} using the selected median values. In the bitrate-specified analysis, only explicitly evaluated bitrate points were retained, and panels with fewer than two available codec series were omitted. In the energy figures, bars show the three-run median and whiskers show the observed minimum and maximum; the whiskers do not represent confidence intervals.

\section{Results and Analysis}

\subsection{Rate--Quality Curves and Overlap Feasibility} \label{sec6_1}

\begin{figure}
    \centering
    \includegraphics[width=1.0\linewidth]{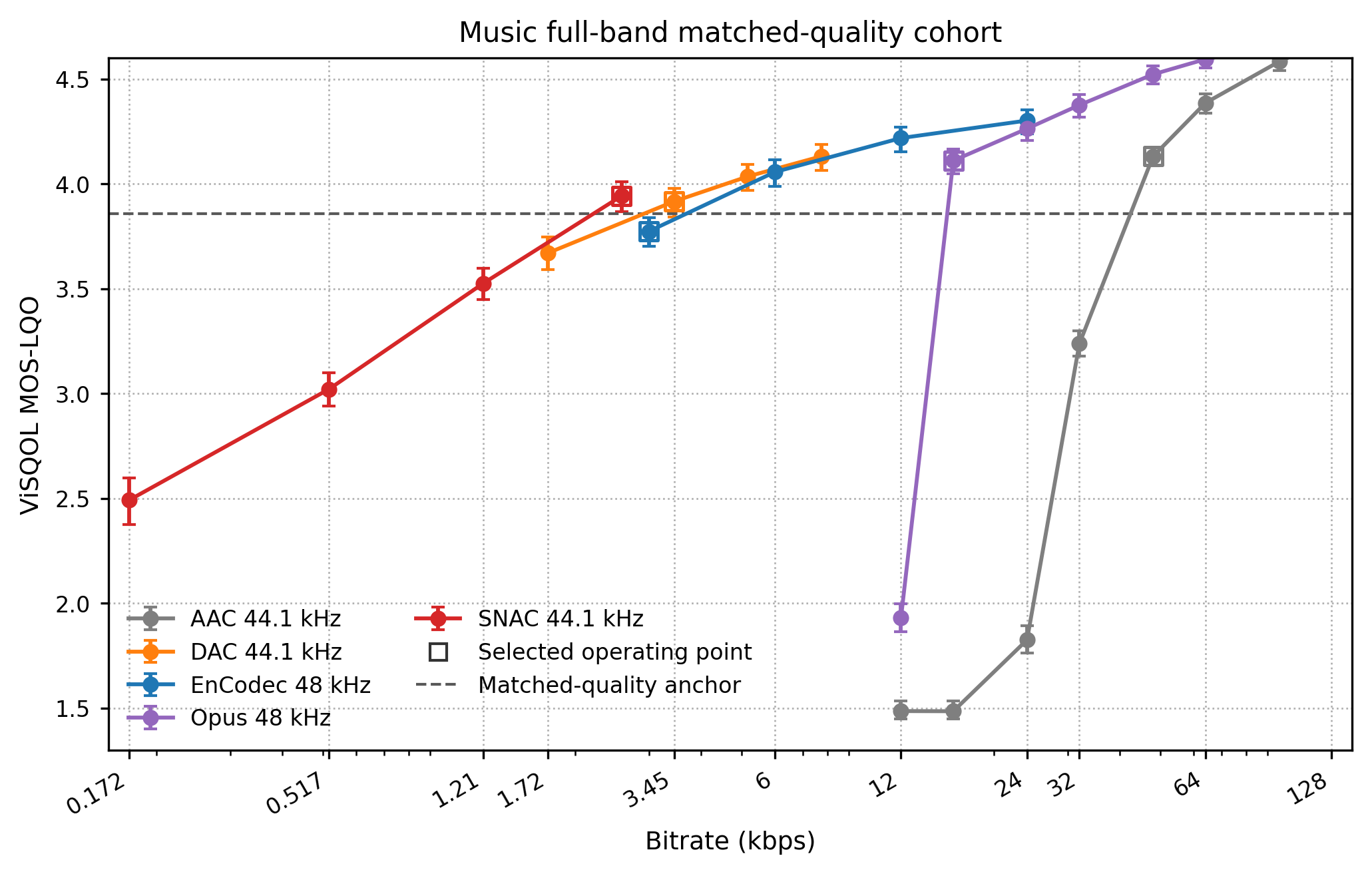}
\caption{Rate--quality curves and overlap-based matched-quality anchor for the full-band-class music cohort. Markers and lines show treatment-level mean ViSQOL values, error bars show 95\% source-track-level bootstrap confidence intervals, open squares indicate the selected measured operating points, and the dashed horizontal line indicates the mid-overlap anchor.}
    \label{fig:1}
\end{figure}
\begin{figure}
    \centering
    \includegraphics[width=1.0\linewidth]{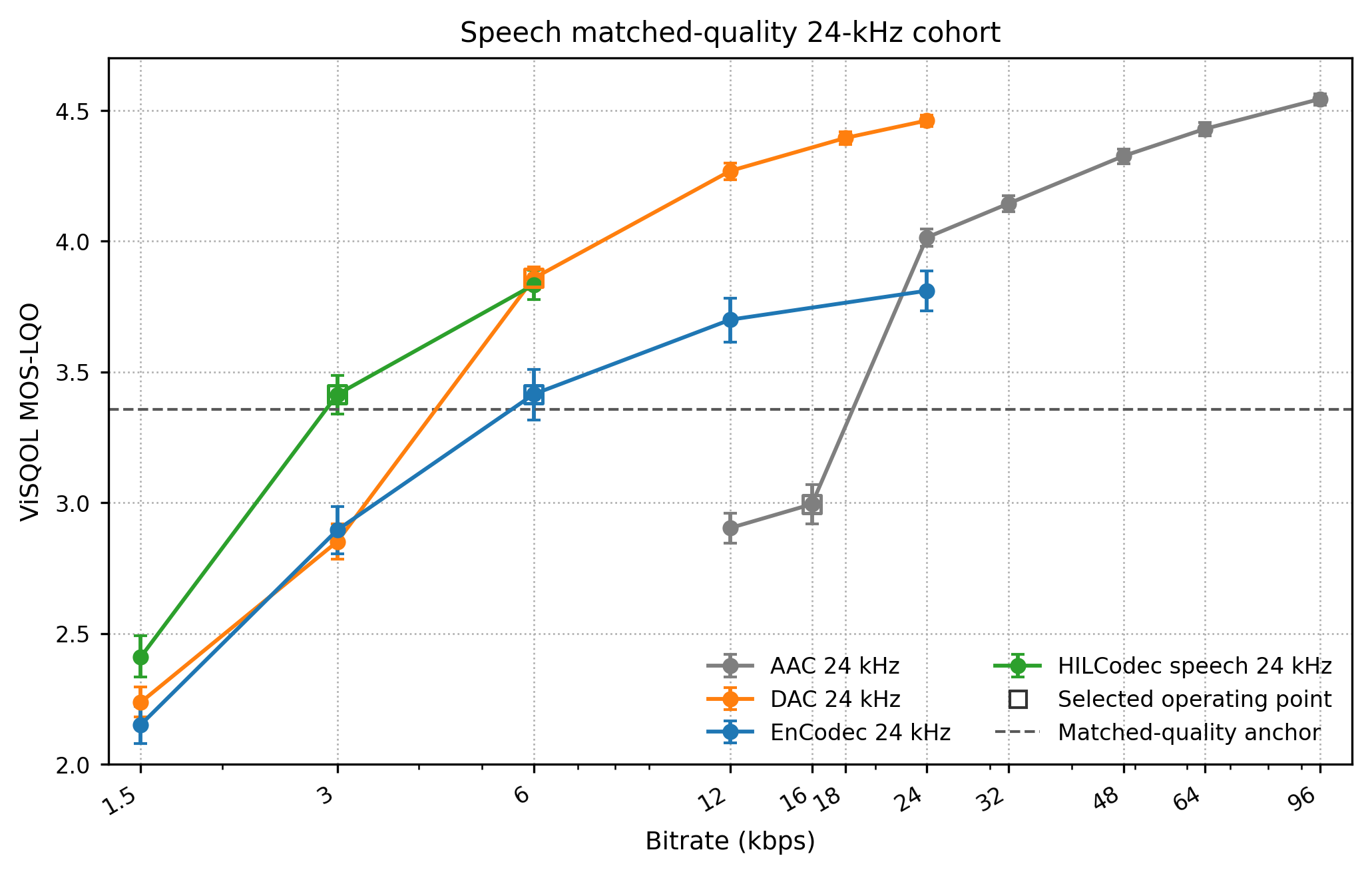}
    \caption{Rate--quality curves and overlap-based matched-quality anchor for the 24-kHz speech cohort. Markers and lines show treatment-level mean ViSQOL values, error bars show 95\% source-track-level bootstrap confidence intervals, open squares indicate the selected measured operating points, and the dashed horizontal line indicates the mid-overlap anchor.}
    \label{fig:2}
\end{figure}

Figs.~\ref{fig:1} and~\ref{fig:2} define the matched-quality anchors used in the energy analysis. Because bitrate ladders and attainable quality ranges differ across codecs, operating points were selected in ViSQOL space rather than by bitrate or codebook count.

Speech and music were evaluated separately under the original-reference protocol. Because the reference was not bandwidth-matched to the codec output, bandwidth loss contributes to the measured quality degradation. The rate--quality curves use FP32 outputs, while precision remains part of the decoder execution configuration.

Fig.~\ref{fig:1} shows the rate--quality curves and overlap-based anchor for the full-band-class music cohort. This cohort includes AAC-LC at 44.1 kHz, Opus at 48 kHz, EnCodec at 48 kHz, DAC at 44.1 kHz, and SNAC at 44.1 kHz. Based on the intersection of the treatment-level mean ViSQOL ranges, the common overlap interval was 3.774--3.943 MOS-LQO. We used the midpoint of this interval, 3.858 MOS-LQO, as the matched-quality anchor for the full-band music cohort. For each codec instance, the operating point closest to this anchor was selected from the measured discrete bitrate ladder.

Fig.~\ref{fig:2} shows the rate--quality curves and overlap-based anchor for the 24-kHz speech cohort. This cohort includes AAC-LC, EnCodec, DAC, and HILCodec speech configurations at 24 kHz. Opus at 24 kHz was not included in this matched-quality cohort because the predefined Opus bitrate ladder evaluated in this study did not contain an operating point within the selected overlap region. SNAC at 24 kHz was also excluded because its attainable quality range was below the common overlap. The resulting overlap interval for this speech cohort was 2.903--3.810 MOS-LQO, and its midpoint, 3.357 MOS-LQO, was used as the speech matched-quality anchor.

The selected operating points were EnCodec at 6~kbps, HILCodec at 3~kbps, DAC at 6~kbps, and AAC-LC at 16~kbps. The DAC point lay above the anchor, whereas the AAC-LC point lay below it.

The selected points are not exactly quality-equivalent. In music, the AAC-LC and Opus points lie above both the anchor and the EnCodec point, which can make the neural bitrate savings appear more favorable than they would in an interpolated exact-quality comparison. In speech, AAC-LC lies below the anchor, whereas the neural points lie at or above it, which produces the opposite bias. The resulting thresholds therefore apply to the selected measured points rather than to exact perceptual equivalence.

\subsection{Quality-Anchor Decoder Energy}
\label{sec6_2}

This subsection compares decoder-side energy at the operating points selected in Section~\ref{sec6_1}. For Laptop, each neural point is represented by the lowest median \(e_{\mathrm{dec}}\) among the evaluated execution-valid paths. For Phone, neural codecs use the XNNPACK CPU path, whereas conventional codecs use the MediaCodec system decoder path. Laptop and Phone values are interpreted separately because their telemetry layers differ.

\begin{figure}
    \centering
    \includegraphics[width=1.0\linewidth]{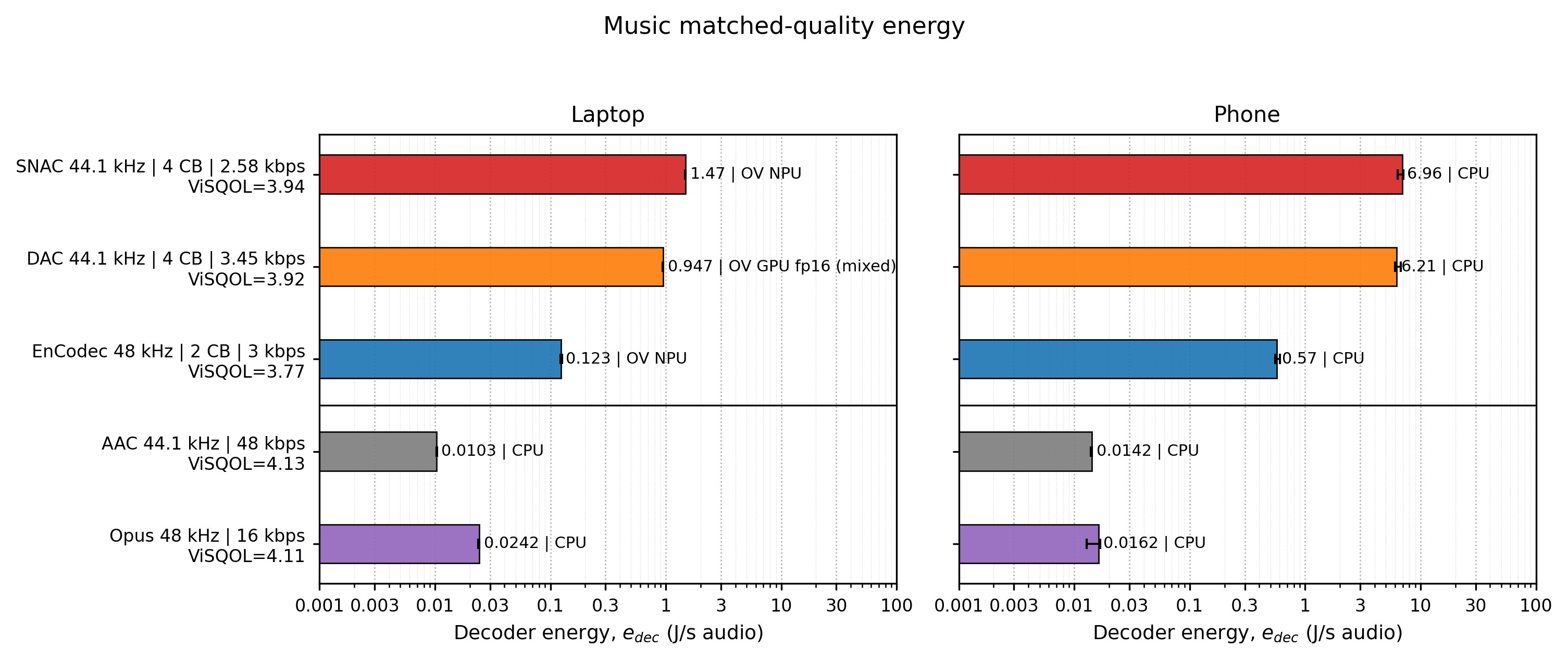}
\caption{Quality-anchor decoder-side energy for the full-band-class music cohort. Laptop values use the lowest-energy execution-valid path evaluated for each neural codec; Phone values use XNNPACK CPU for neural codecs and the MediaCodec system path for conventional codecs. Bars show the median of three repeated runs, and whiskers show the observed minimum and maximum.}
    \label{fig:3}
\end{figure}

Fig.~\ref{fig:3} shows the quality-anchor decoder-energy results for the full-band-class music cohort. EnCodec, DAC, and SNAC were selected at 3, 3.45, and 2.584~kbps, respectively. Opus and AAC-LC were selected at 16 and 48~kbps. The selected AAC-LC and Opus points have higher ViSQOL scores than the selected EnCodec point because selection was restricted to the measured discrete ladders.

On Laptop, the three-run median \(e_{\mathrm{dec}}\) was 0.123~J/s audio for EnCodec on OpenVINO NPU, 0.947~J/s for DAC on the OpenVINO GPU FP16-requested path, and 1.47~J/s for SNAC on OpenVINO NPU. The corresponding AAC-LC and Opus medians on the native CPU decoder path were 0.0103 and 0.0242~J/s, respectively. Because the DAC path included approximately 45\% non-target execution, it is interpreted as a mixed GPU-requested path rather than as pure GPU execution.

\begin{figure}
    \centering
    \includegraphics[width=1.0\linewidth]{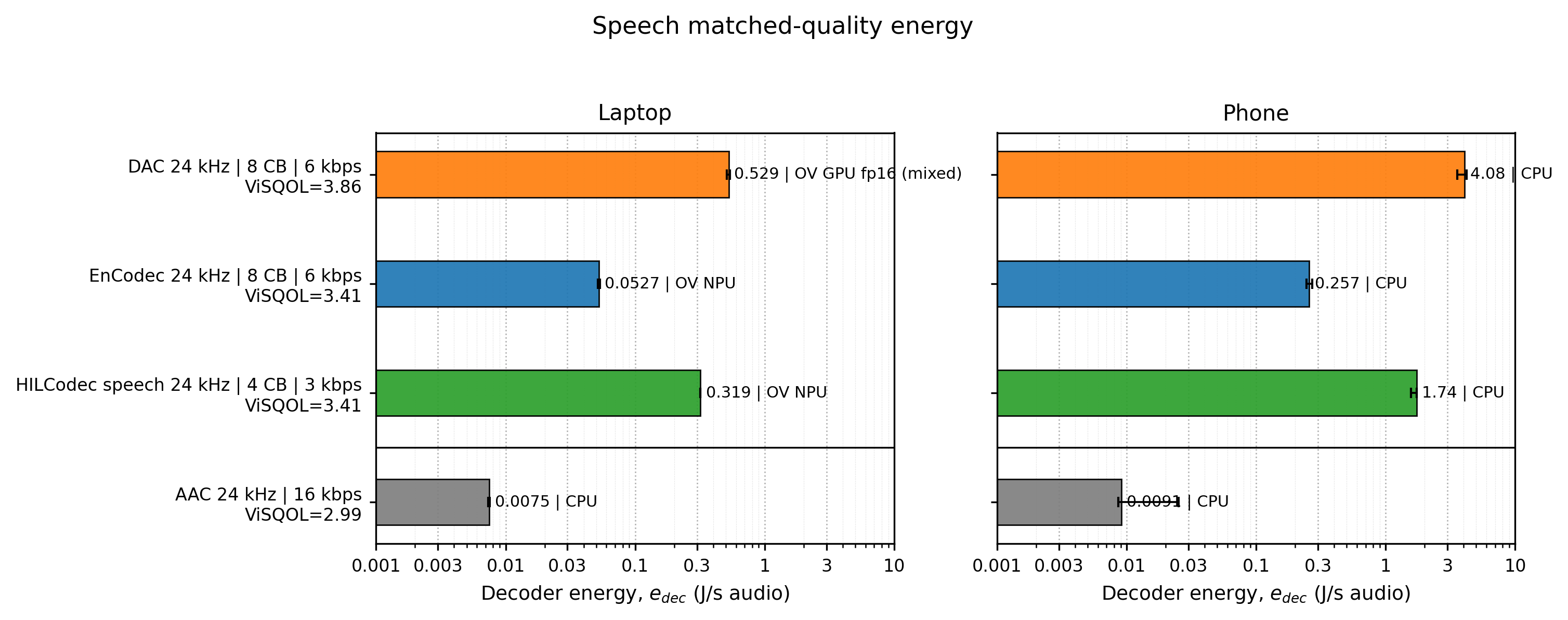}
\caption{Quality-anchor decoder-side energy for the 24-kHz speech cohort. Laptop values use the lowest-energy execution-valid path evaluated for each neural codec; the conventional value uses the native CPU decoder path. Phone values use XNNPACK CPU for neural codecs and the MediaCodec system path for the conventional codec. Bars show the median of three repeated runs, and whiskers show the observed minimum and maximum.}
    \label{fig:4}
\end{figure}

On Phone, the median \(e_{\mathrm{dec}}\) values were 0.570~J/s for EnCodec, 6.21~J/s for DAC, and 6.96~J/s for SNAC on XNNPACK CPU, compared with 0.0142~J/s for AAC-LC and 0.0162~J/s for Opus on the MediaCodec system path. Thus, each selected neural path required more decoder-side energy than either conventional path within the Phone measurement layer. Fig.~\ref{fig:4} shows the corresponding results for the 24-kHz speech cohort.

 On Laptop, the median \(e_{\mathrm{dec}}\) values were 0.0527~J/s audio for EnCodec 24~kHz at 6~kbps on OpenVINO NPU, 0.319~J/s for HILCodec speech 24~kHz at 3~kbps on OpenVINO NPU, and 0.529~J/s for DAC 24~kHz at 6~kbps on the OpenVINO GPU FP16-requested path. AAC-LC 24~kHz at 16~kbps required 0.00750~J/s on the native CPU path. The selected DAC path contained approximately 47\% non-target execution and is therefore interpreted as mixed execution.

On Phone, the medians were 0.257~J/s for EnCodec, 1.74~J/s for HILCodec, and 4.08~J/s for DAC on XNNPACK CPU, compared with 0.00909~J/s for AAC-LC on the system decoder path. EnCodec had the lowest decoder energy among the selected neural speech codecs, whereas HILCodec provided the lowest neural bitrate. The observed AAC-LC range on Phone was 0.00866--0.0251 J/s. The median is retained as the representative value, while the wider range remains visible in the figure without being interpreted as a confidence interval.

\subsection{Decoder Energy at Specified Bitrate Settings}
\label{sec6_3}

The bitrate-specified comparison reports ViSQOL and decoder-side energy for operating points available at predefined bitrate settings. Unlike the matched-quality comparison, this analysis does not align codecs by perceptual quality.

\begin{figure}[t]
    \centering
    \includegraphics[width=1.0\linewidth]{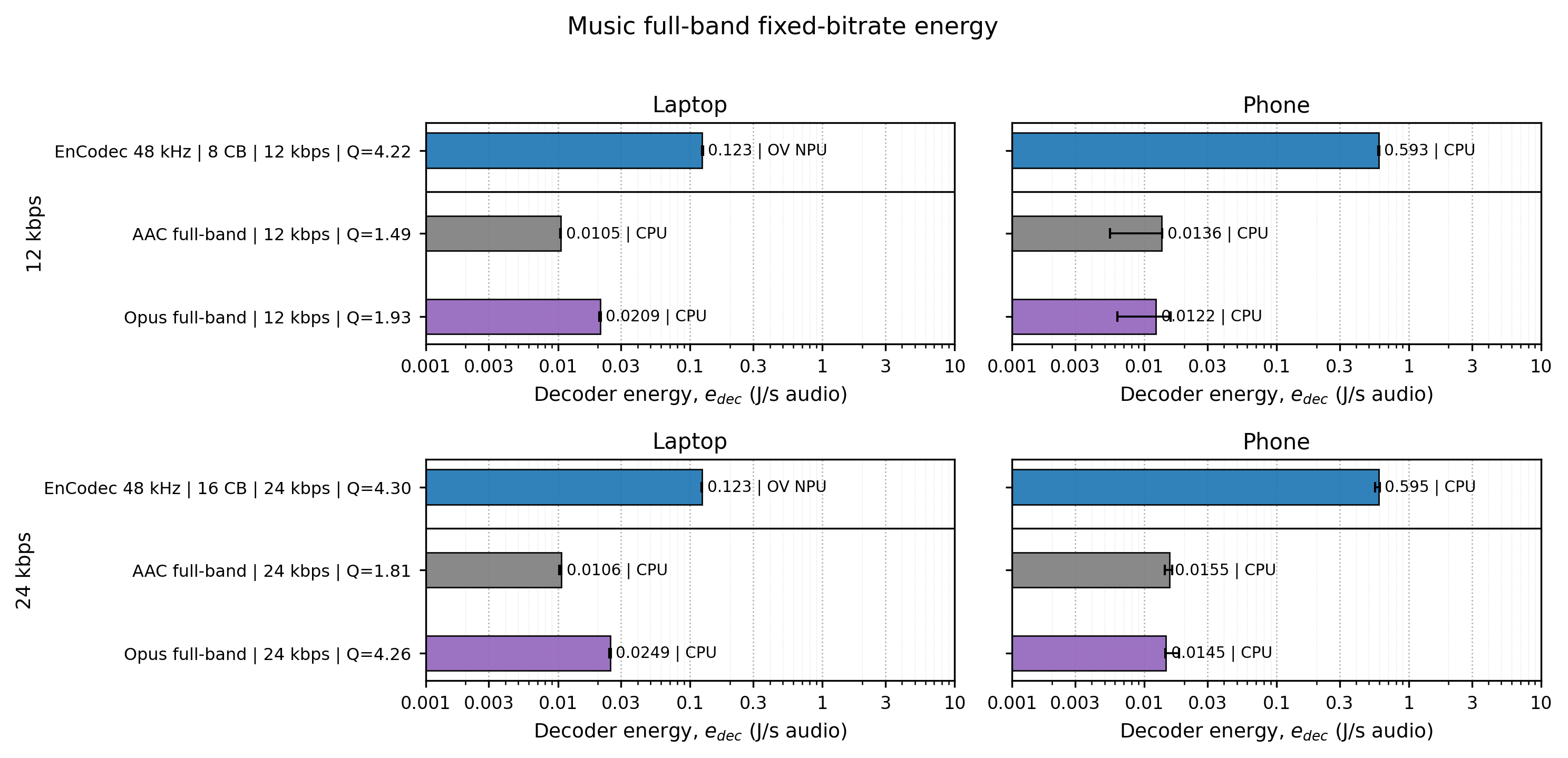}
    \caption{Bitrate-specified decoder-side energy and ViSQOL for the full-band-class music setting. Only explicitly evaluated operating points are included. Panels with fewer than two available codec series are omitted, leaving the 12- and 24-kbps conditions. Energy bars show the three-run median and whiskers show the observed minimum and maximum.}
    \label{fig:5}
\end{figure}

Fig.~\ref{fig:5} reports the full-band-class music conditions for which at least two codec series were explicitly evaluated at the same requested bitrate. The 3 and 6~kbps conditions are omitted because only EnCodec was available; no cross-codec inference is made at these two bitrates.

At 12~kbps, EnCodec achieved a ViSQOL score of 4.22, compared with 1.49 for AAC-LC and 1.93 for Opus under the original-reference full-band protocol. On Laptop, EnCodec, AAC-LC, and Opus required 0.123, 0.0105, and 0.0209~J/s audio, respectively. On Phone, the corresponding values were 0.593, 0.0136, and 0.0122~J/s audio. At 24~kbps, EnCodec and Opus achieved similar ViSQOL scores of 4.30 and 4.26, respectively, while AAC-LC scored 1.81. On Laptop, Opus and EnCodec required 0.0249 and 0.123~J/s audio, respectively. On Phone, the corresponding values were 0.0145 and 0.595~J/s audio. Under this protocol, EnCodec achieved higher ViSQOL at 12~kbps. At 24~kbps, Opus achieved similar ViSQOL with lower decoder energy.

\begin{figure}
    \centering
    \includegraphics[width=1.0\linewidth]{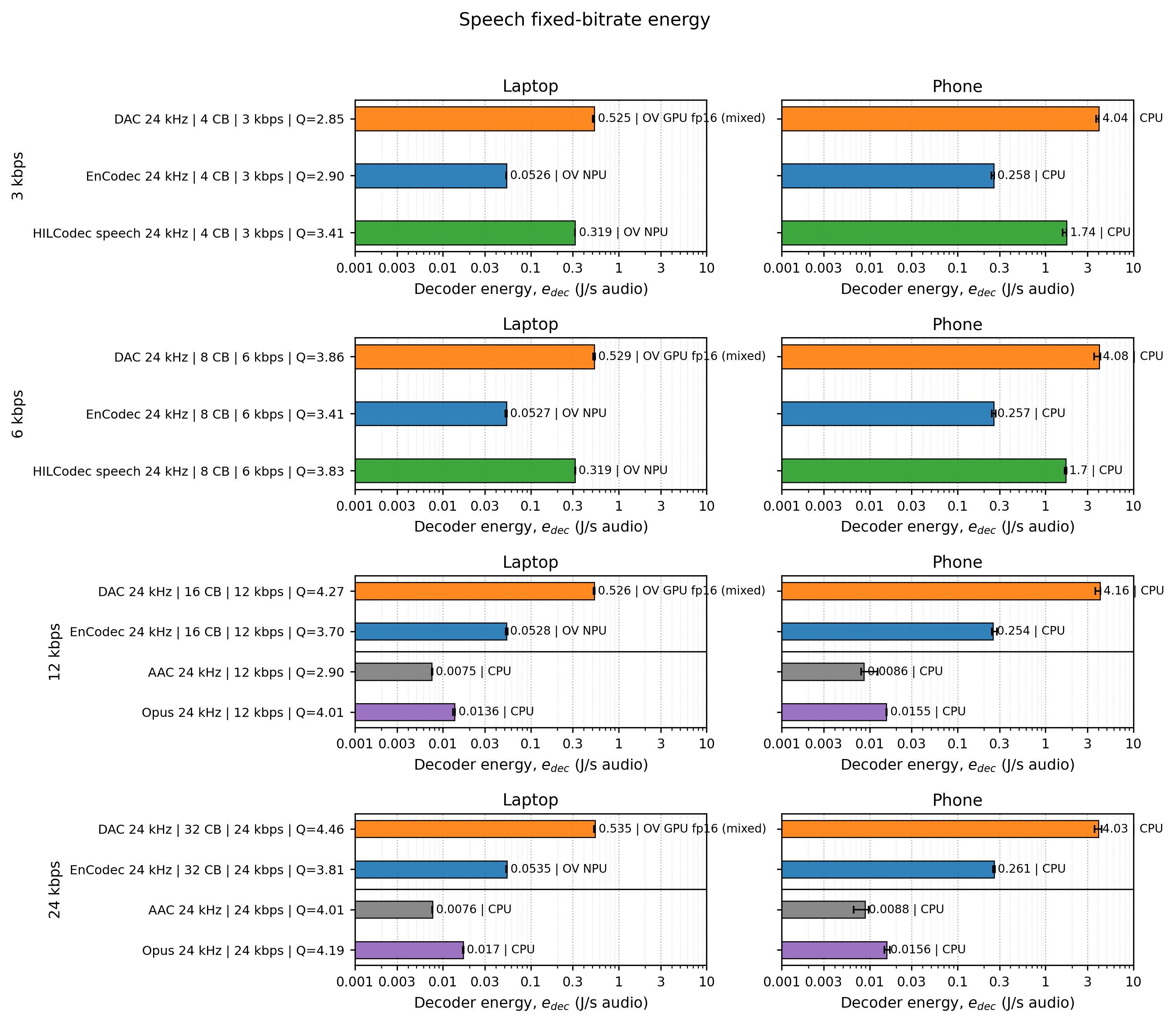}
    \caption{Bitrate-specified decoder-side energy and ViSQOL for the 24-kHz speech setting at 3, 6, 12, and 24~kbps. Only explicitly evaluated operating points are included. Energy bars show the three-run median and whiskers show the observed minimum and maximum.}
    \label{fig:6}
\end{figure}

Fig.~\ref{fig:6} shows the 24-kHz speech results. At 3~kbps, HILCodec achieved the highest ViSQOL score, 3.41; EnCodec and DAC scored 2.90 and 2.85, respectively. On Laptop, EnCodec, HILCodec, and DAC required 0.0526, 0.319, and 0.525~J/s audio, respectively. On Phone, the corresponding values were 0.258, 1.74, and 4.04~J/s audio. At 6~kbps, HILCodec and DAC achieved scores of 3.83 and 3.86, respectively, compared with 3.41 for EnCodec. Despite its lower ViSQOL score, EnCodec again had the lowest neural decoder energy, requiring 0.0527 J/s audio on Laptop and 0.257 J/s audio on Phone.

At 12 kbps, Opus achieved higher ViSQOL than EnCodec (4.01 versus 3.70) while requiring less decoder energy on both platforms. DAC achieved the highest ViSQOL score, 4.27, but required 0.526 J/s audio on Laptop and 4.16 J/s audio on Phone. At 24 kbps, AAC-LC and Opus continued to require relatively low decoder energy, whereas DAC achieved the highest ViSQOL score, 4.46, at the cost of higher decoder energy.

Across the evaluated neural operating points, decoder energy varied less with codebook depth than the ViSQOL scores. This trend is specific to the evaluated runtime configurations and should not be assumed to generalize to other decoder architectures.

\subsection{Parametric Transmission-Energy Break-Even Analysis}
\label{sec6_4}

The matched-quality results in Figs.~\ref{fig:3} and \ref{fig:4} show lower bitrates but higher decoder-side energy for the selected neural points relative to the applicable conventional references. Music points are compared separately with AAC-LC and Opus, whereas speech points are compared only with AAC-LC because Opus was not part of the matched-quality speech cohort. Each \(\alpha^*_{A,B}\) was calculated from Eq.~\eqref{eq:break_even} using the selected median \(e_{\mathrm{dec}}\) values. A lower threshold indicates that the neural point becomes favorable at a lower modeled \(\alpha_N\). The thresholds are model-derived quantities rather than measurements of a particular wireless interface. Values above 200~mJ/kbit are annotated with their exact values at the upper plotting limit.

\begin{figure}
    \centering
    \includegraphics[width=1.0\linewidth]{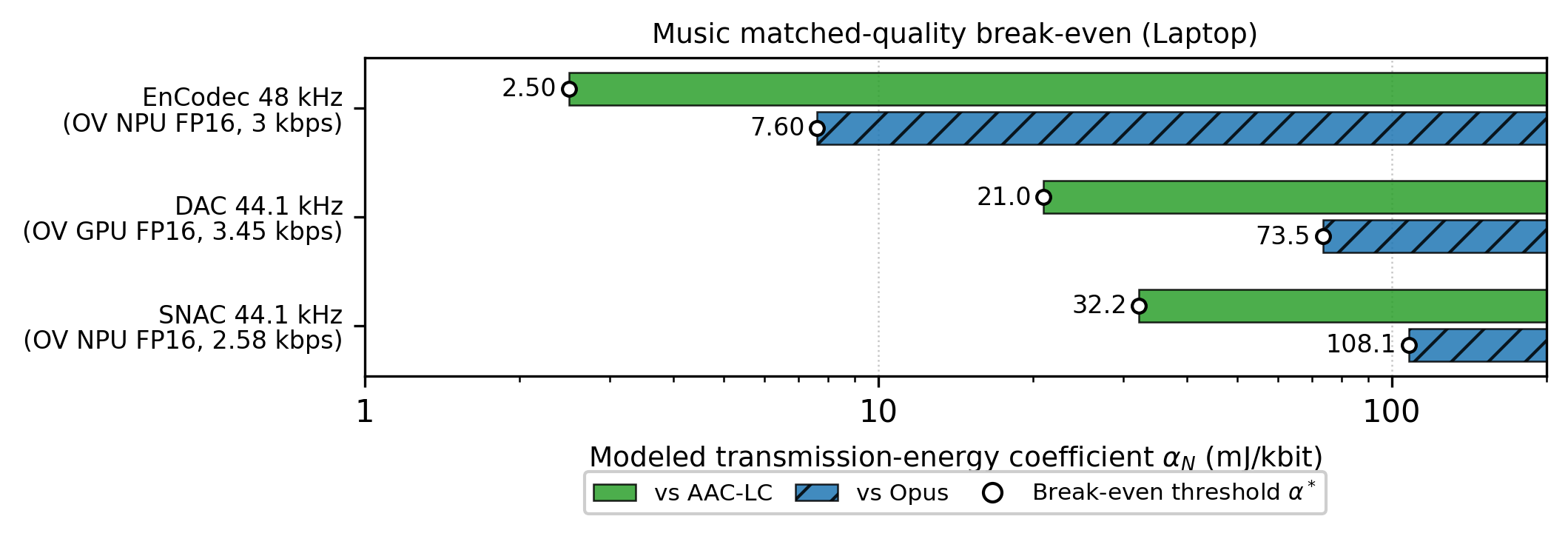}
\caption{Pairwise break-even thresholds relative to AAC-LC and Opus for the full-band-class music cohort on Laptop. Markers denote \(\alpha^*\); rightward intervals denote the neural-favorable region \(\alpha_N>\alpha^*\).}
    \label{fig:7}
\end{figure}

Fig.~\ref{fig:7} shows the analytic music thresholds on Laptop. EnCodec reached break-even at 2.50~mJ/kbit relative to AAC-LC and 7.60~mJ/kbit relative to Opus. The corresponding DAC thresholds were 21.0 and 73.5~mJ/kbit, and the SNAC thresholds were 32.2 and 108.1~mJ/kbit. EnCodec had the lowest threshold against both baselines. The DAC values represent its lowest-energy execution-valid GPU-requested path, which included partial non-target execution.

\begin{figure}
    \centering
    \includegraphics[width=1.0\linewidth]{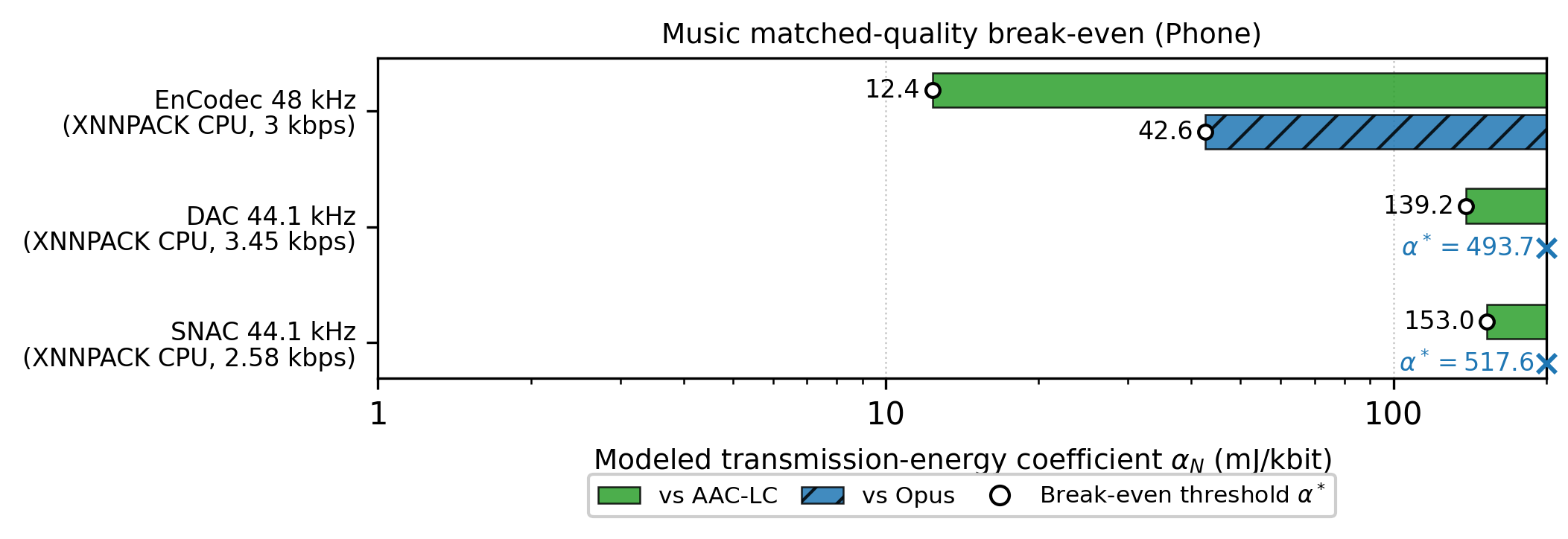}
\caption{Pairwise break-even thresholds relative to AAC-LC and Opus for the full-band-class music cohort on Phone. Markers denote \(\alpha^*\); rightward intervals denote the neural-favorable region \(\alpha_N>\alpha^*\). Thresholds above 200~mJ/kbit are annotated with their exact values at the upper plotting boundary.}
    \label{fig:8}
\end{figure}

Fig.~\ref{fig:8} shows the music thresholds on Phone. EnCodec reached break-even at 12.4~mJ/kbit relative to AAC-LC and 42.6~mJ/kbit relative to Opus. DAC reached 139.2~mJ/kbit relative to AAC-LC and 493.7~mJ/kbit relative to Opus. SNAC reached 153.0 and 517.6~mJ/kbit, respectively. The two Opus-referenced thresholds exceed the 200~mJ/kbit plotting range and are annotated with their exact values at the plot boundary in Fig.~\ref{fig:8}.

\begin{figure}
    \centering
    \includegraphics[width=1.0\linewidth]{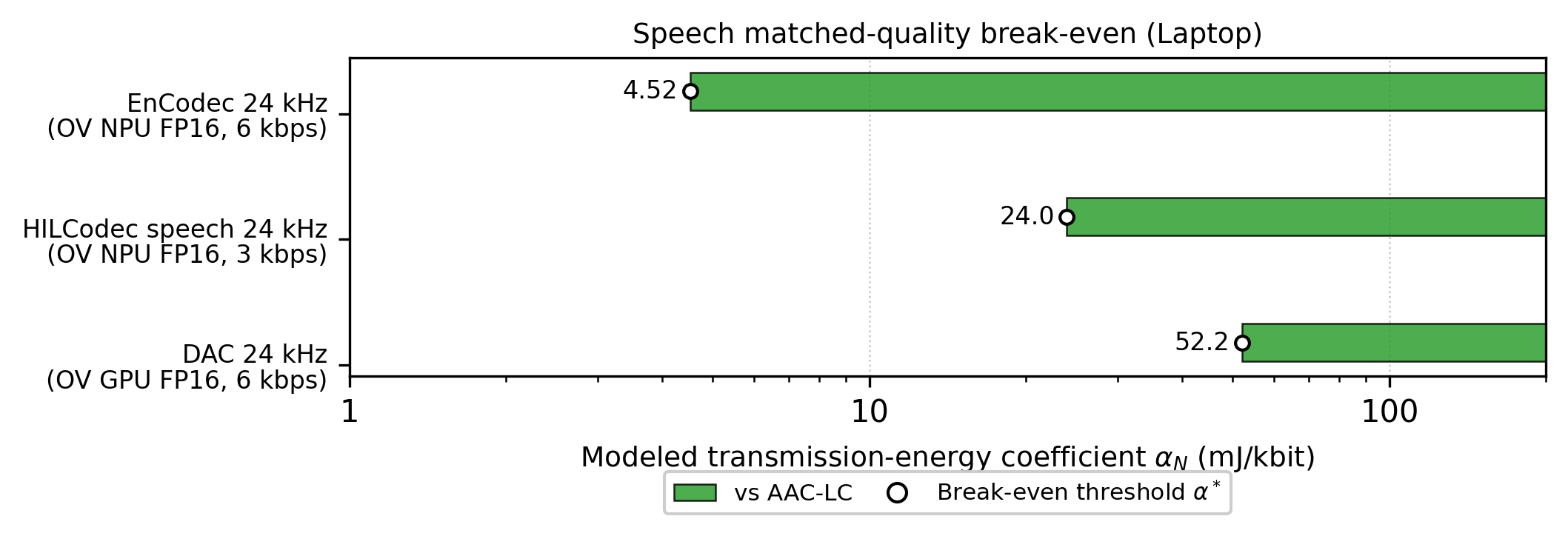}
\caption{Pairwise break-even thresholds relative to AAC-LC for the 24-kHz speech cohort on Laptop. Markers denote \(\alpha^*\); rightward intervals denote the neural-favorable region \(\alpha_N>\alpha^*\).}
    \label{fig:9}
\end{figure}

\begin{figure}
    \centering
    \includegraphics[width=1.0\linewidth]{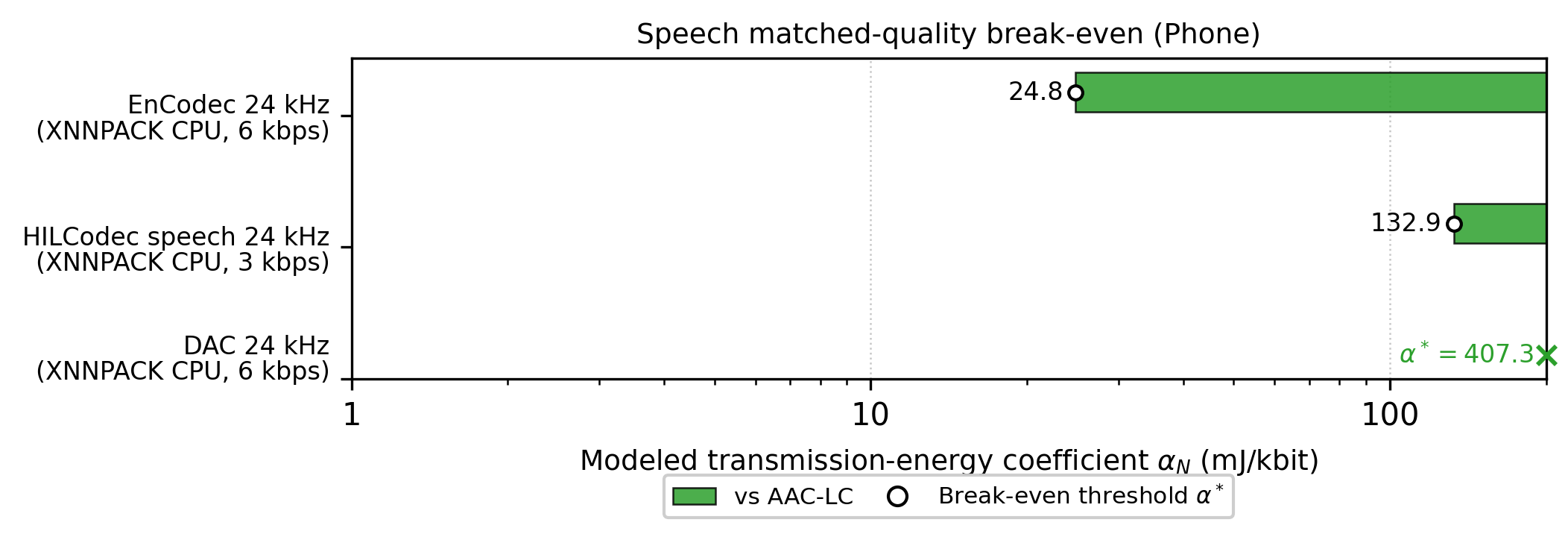}
\caption{Pairwise break-even thresholds relative to AAC-LC for the 24-kHz speech cohort on Phone. Markers denote \(\alpha^*\); rightward intervals denote the neural-favorable region \(\alpha_N>\alpha^*\). Thresholds above 200~mJ/kbit are annotated with their exact values at the upper plotting boundary.}
    \label{fig:10}
\end{figure}

Figs.~\ref{fig:9} and~\ref{fig:10} show the AAC-LC-referenced thresholds for the 24-kHz speech cohort. On Laptop, EnCodec, HILCodec, and DAC reached break-even at 4.52, 24.0, and 52.2~mJ/kbit, respectively. On Phone, their thresholds were 24.8, 132.9, and 407.3~mJ/kbit, respectively. The Phone DAC threshold of 407.3~mJ/kbit exceeds the 200~mJ/kbit plotting range and is annotated with its exact value at the plot boundary in Fig.~\ref{fig:10}. EnCodec had the lowest threshold among the evaluated neural speech points on both platforms.

The lower AAC-LC-referenced music thresholds reflect the larger bitrate reductions relative to the selected AAC-LC point at 48~kbps compared to the Opus point at 16~kbps. Their exact values also depend on the corresponding decoder-energy differences.

The speech analysis has narrower baseline coverage than the music analysis because Opus did not have an evaluated matched-quality point in the selected overlap cohort. These values are AAC-LC-referenced results for the measured discrete points and do not provide a general ranking against all conventional speech codecs.

\subsection{Execution Validity and Latency Diagnostics}
\label{sec6_5}

Requested and effective execution paths can differ; therefore, runtime failures, provider coverage, and non-target execution were recorded separately. Paths that failed during session creation or decoding, or showed no effective use of the requested accelerator, were excluded from the main comparison.

\begin{figure}
    \centering
    \includegraphics[width=1.0\linewidth]{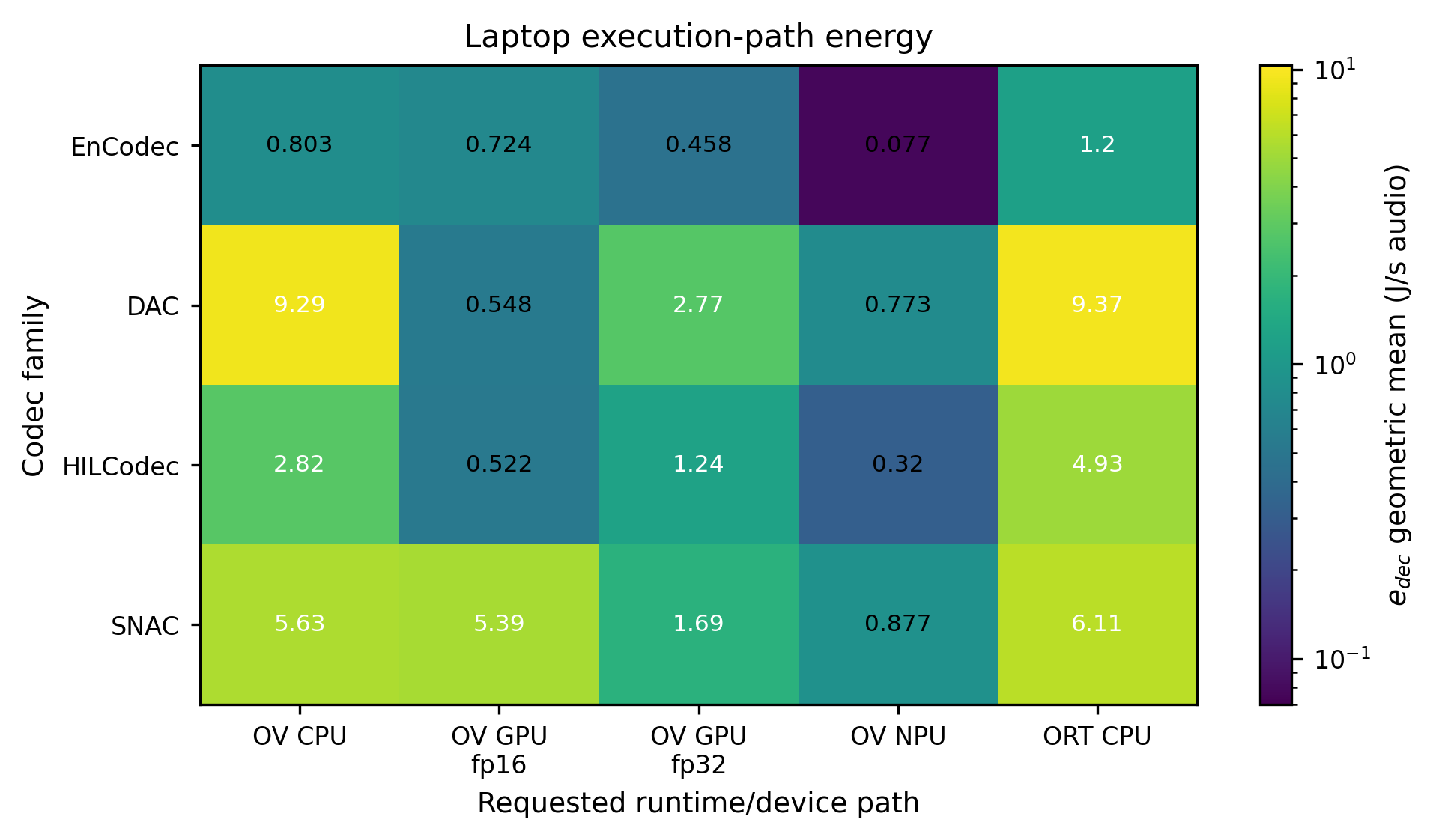}
    \caption{Family-level geometric mean of per-operating-point median decoder energy across requested Laptop runtime and device paths. Each family is summarized over operating points successfully measured on all compared paths.}
    \label{fig:11}
\end{figure}

Fig.~\ref{fig:11} summarizes the geometric mean of the per-operating-point median decoder energy over the complete set of operating points successfully measured on all compared Laptop paths within each codec family. Among these common operating points, the NPU path produced the lowest family-level geometric-mean decoder energy for EnCodec, HILCodec, and SNAC. For DAC, the GPU FP16-requested path produced the lowest value but included substantial non-target execution.

\begin{figure}
    \centering
    \includegraphics[width=1.0\linewidth]{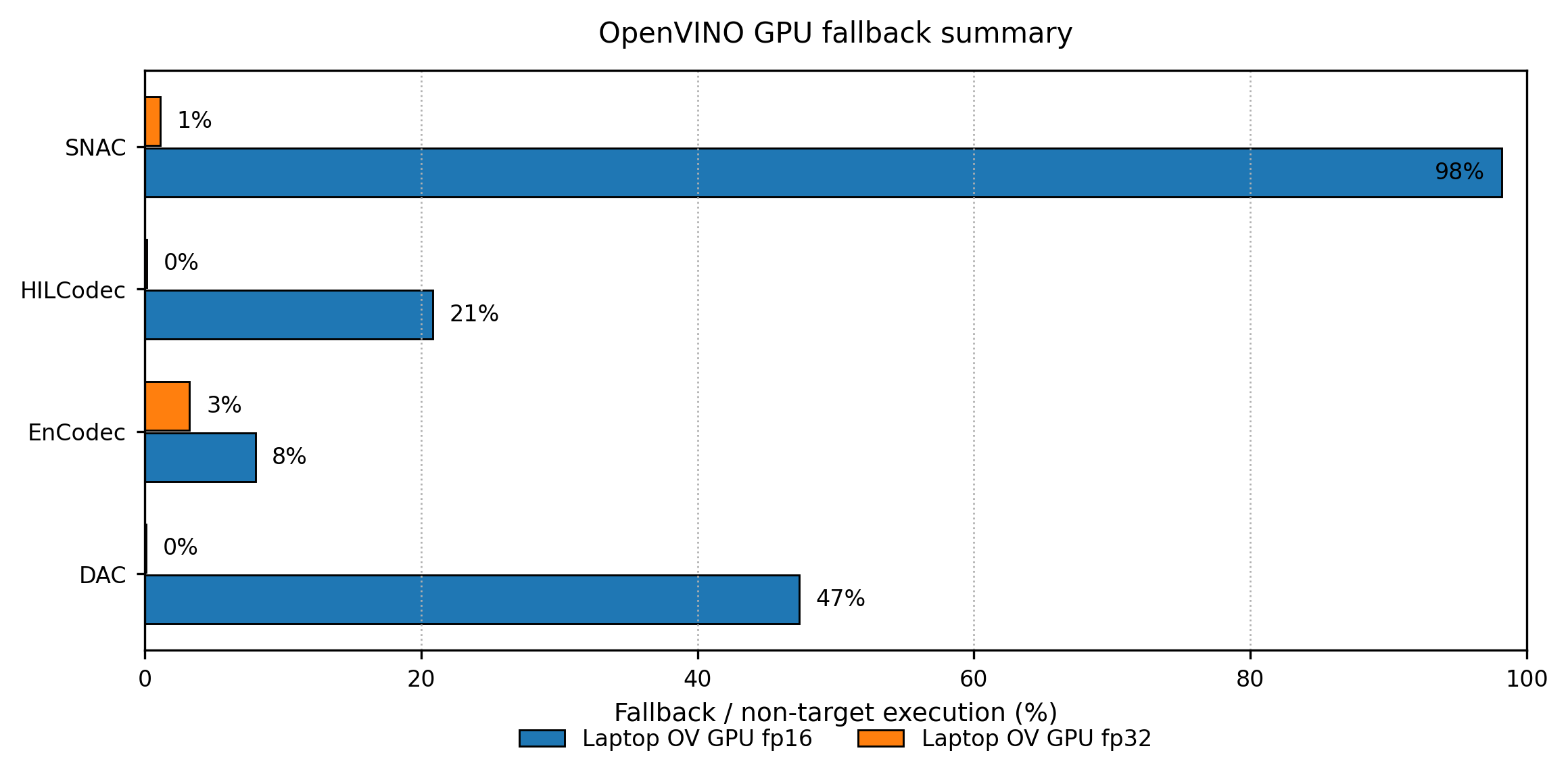}
    \caption{OpenVINO GPU fallback summary for neural codecs. Values are medians across the evaluated operating points within each codec family and requested GPU precision. Fallback ratio is the share of profiled execution time assigned to non-target devices.}
    \label{fig:12}
\end{figure}

Fig.~\ref{fig:12} shows the OpenVINO GPU fallback summary. For the FP16-requested path, the non-target shares were 98\% for SNAC, 47\% for DAC, 21\% for HILCodec, and 8\% for EnCodec. DAC nevertheless had its lowest measured energy on this mixed path. Partial non-target execution was reported as mixed execution rather than treated as failure; paths with no effective target use were excluded.

Latency was used only as a feasibility diagnostic. All selected Laptop paths required less than 1~s to decode 1~s of audio. On Phone, all selected paths except full-band DAC and SNAC also met this condition; these two paths required 1.32 and 1.63~s, respectively, on XNNPACK CPU.

\begin{table}[t]
\centering
\caption{Latency for decoding 1 s of audio at selected operating points.}
\label{tab:decoder_latency}
\scriptsize
\setlength{\tabcolsep}{2.4pt}
\renewcommand{\arraystretch}{1.08}
\begin{threeparttable}
\begin{tabularx}{\columnwidth}{@{}
c
>{\raggedright\arraybackslash}p{0.36\columnwidth}
>{\raggedright\arraybackslash}X
>{\raggedright\arraybackslash}X
@{}}
\toprule
\textbf{D}
& \textbf{Codec / operating point}
& \textbf{Laptop path / latency}
& \textbf{Phone path / latency} \\
\midrule

M & EnCodec 48 kHz, 2CB, 3 kbps
& \makecell[l]{OV-NPU FP16\\19.1 ms}
& \makecell[l]{XNNPACK CPU\\118 ms} \\

M & DAC 44.1 kHz, 4CB, 3.45 kbps
& \makecell[l]{OV-GPU FP16 req.\\33.1 ms}
& \makecell[l]{XNNPACK CPU\\1316 ms} \\

M & SNAC 44.1 kHz, 4CB, 2.58 kbps
& \makecell[l]{OV-NPU FP16\\161 ms}
& \makecell[l]{XNNPACK CPU\\1632 ms} \\

M & AAC-LC 44.1 kHz, 48 kbps
& \makecell[l]{FFmpeg CPU\\0.80 ms}
& \makecell[l]{MediaCodec system\\74.3 ms} \\

M & Opus 48 kHz, 16 kbps
& \makecell[l]{FFmpeg CPU\\1.63 ms}
& \makecell[l]{MediaCodec system\\46.2 ms} \\

\midrule

S & EnCodec 24 kHz, 8CB, 6 kbps
& \makecell[l]{OV-NPU FP16\\8.79 ms}
& \makecell[l]{XNNPACK CPU\\53.4 ms} \\

S & HILCodec 24 kHz, 4CB, 3 kbps
& \makecell[l]{OV-NPU FP16\\40.8 ms}
& \makecell[l]{XNNPACK CPU\\307 ms} \\

S & DAC 24 kHz, 8CB, 6 kbps
& \makecell[l]{OV-GPU FP16 req.\\18.4 ms}
& \makecell[l]{XNNPACK CPU\\748 ms} \\

S & AAC-LC 24 kHz, 16 kbps
& \makecell[l]{FFmpeg CPU\\0.60 ms}
& \makecell[l]{MediaCodec system\\52.1 ms} \\

\bottomrule
\end{tabularx}
\begin{tablenotes}[flushleft]
\footnotesize
\item D denotes the domain: M for music and S for speech. OV denotes OpenVINO. CB denotes codebooks. Latency is the median wall-clock time required to decode 1 s of audio.
\end{tablenotes}
\end{threeparttable}
\end{table}

\section{Discussion}

At the matched-quality points, each selected neural operating point had a lower bitrate but required more decoder-side energy than the applicable conventional points within each platform measurement layer. EnCodec yielded the lowest neural break-even thresholds in both cohorts on both platforms. DAC and SNAC required substantially larger transmission-energy coefficients on the Phone XNNPACK CPU path. Their full-band Phone configurations also required more than 1~s to decode 1~s of audio, so the corresponding steady-state energy values do not demonstrate real-time playback feasibility on that path.

The execution-path results show that a requested accelerator label alone is insufficient to characterize actual deployment behavior. EnCodec, HILCodec, and SNAC achieved their lowest Laptop medians on the evaluated NPU paths. DAC achieved its lowest median on an OpenVINO GPU FP16-requested path with substantial non-target execution; therefore, this result represents mixed execution rather than pure GPU execution. The DirectML path did not satisfy the validity criteria, and the Phone QNN path showed no effective use of the requested accelerator. These outcomes limit conclusions about alternative accelerator mappings, particularly for accelerator-assisted Phone deployment. The Phone measurements specifically characterize XNNPACK CPU execution and therefore do not establish the energy behavior of mobile accelerator paths.

The bitrate-specified analysis examines the operating space without assuming quality equivalence. At 12 kbps, EnCodec achieved higher full-band ViSQOL than the evaluated AAC-LC and Opus points under the study protocol. At 24 kbps, Opus reached a similar ViSQOL region with substantially lower decoder energy. For speech, the neural codecs provided operating points in the 3–6-kbps range. AAC-LC and Opus remained strong quality–energy baselines at 12–24 kbps. Overall, the observed rate–energy–quality trade-off depends on both the service bitrate range and the decoder path available on the client.

The matched-quality analysis is constrained by discrete bitrate ladders. In the music cohort, the selected AAC-LC and Opus points lie above the anchor and above the selected EnCodec point, which can make the neural bitrate savings and corresponding break-even thresholds appear more favorable than in an exact continuously matched comparison. 
In the speech cohort, the selected AAC-LC point lies below the anchor, whereas the neural points lie at or above it, which produces the opposite mismatch. The reported thresholds are pairwise results for the measured points and should not be interpreted as universal codec rankings. The study protocol also retains bandwidth loss as part of the measured quality degradation rather than bandwidth-matching the reference.

Several limitations define the scope of these results. Measurements were obtained from one Laptop and one Phone, so realized energy may vary with model conversion, SoC, driver, runtime version, or accelerator support. The different telemetry layers also restrict absolute energy comparisons to within each platform: Laptop reports processor package-level energy, whereas Phone reports device-level battery-discharge energy. Three repetitions support reporting of the median and observed range but do not characterize the full run-to-run distribution. The ViSQOL anchors represent objective-metric operating points rather than exact subjective equivalence, and objective and subjective codec rankings can differ across metrics and codec conditions ~\cite{lanzendorfer2025evaluating,mack2026assessing}.

The evaluated codec set is limited to publicly available decoders that could be represented and validated within the common benchmark and therefore excludes closed or non-exportable systems. The SNAC lower-rate configurations are retained-scale-derived operating points rather than separately trained, officially released low-bitrate checkpoints. Opus did not provide a matched-quality speech point within the evaluated ladder, and server-side encoding energy remains outside the one-to-many playback model.

These measurements do not establish a codec-wide ranking. A lower bitrate can reduce modeled transmission-plus-decoding energy only when the effective decoder path is sufficiently efficient, and the transmission-energy coefficient exceeds the corresponding pairwise threshold. Decoder architecture, runtime and device mapping, execution validity, and accelerator availability must therefore be considered together with rate–quality performance.

\section{Conclusion}

This paper presented a measurement-based rate–energy–quality analysis of four neural audio codecs and two conventional baselines on Laptop and Phone. Speech and music were evaluated separately under the original-reference ViSQOL protocol, and the main comparison used the measured operating points nearest to the midpoint of the common treatment-level mean ViSQOL overlap. The secondary analysis retained only explicitly evaluated bitrate settings. Decoder energy was reported as idle-subtracted J/s audio, and only execution-valid paths were included in the main comparison. Pairwise break-even thresholds were derived analytically from the measured decoder energy and bitrate.

Across the two matched-quality cohorts, the selected neural operating points achieved similar ViSQOL scores at lower bitrates but required more decoder-side energy than the corresponding conventional paths. EnCodec yielded the lowest neural break-even thresholds in both cohorts. In contrast, DAC and SNAC on the Phone XNNPACK CPU path produced the highest thresholds, with several comparisons exceeding 200 mJ/kbit; their full-band configurations also required more than 1 s to decode 1 s of audio. The Laptop results further showed that the lowest measured neural energy depended strongly on runtime and target mapping. In particular, the DAC GPU FP16-requested result contained substantial non-target execution and should therefore be interpreted as mixed execution rather than pure GPU decoding.

A low bitrate alone is therefore insufficient to establish an energy benefit for client deployment. Within the modeled transmission-plus-decoding boundary, a neural operating point becomes favorable only when its decoder cost is sufficiently low on the effective execution path, and the transmission-energy coefficient exceeds the corresponding pairwise threshold. The failed DirectML path and the lack of a valid Phone QNN path limit conclusions about alternative accelerator-assisted deployment. Future work should extend these measurements to additional devices, runtime versions, and valid mobile accelerator paths; increase the number of repeated energy measurements; and complement the objective-quality anchors with subjective listening tests.

\section*{Acknowledgments}
This work was supported by the National Research Foundation of Korea(NRF) grant
funded by the Korea government(MSIT) (RS-2025-23523113)
The authors used ChatGPT and Claude to assist with benchmark-script implementation, debugging, and language editing. All AI-assisted outputs were reviewed and revised by the authors as needed. The authors remain responsible for the study design, experiments, analysis, interpretation, and final manuscript.

\bibliographystyle{IEEEtran}
\bibliography{references}

\begin{IEEEbiography}[{\includegraphics[width=1in,height=1.25in,clip,keepaspectratio]{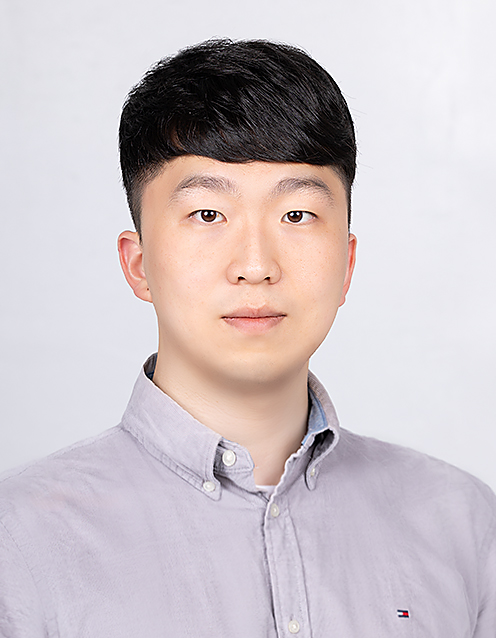}}]{SEUNGHYEON SHIN} received the B.S. degree in mechanical engineering from Dong-A University, Busan, in 2020, and the M.S. degree in electronic engineering from Kyungpook National University, Daegu, in 2022. He is currently a Ph.D. student at Kyungpook National University. His research interests include acoustic anomaly detection, acoustic source separation, and acoustic signal processing.
\end{IEEEbiography}

\begin{IEEEbiography}[{\includegraphics[width=1in,height=1.25in,clip,keepaspectratio]{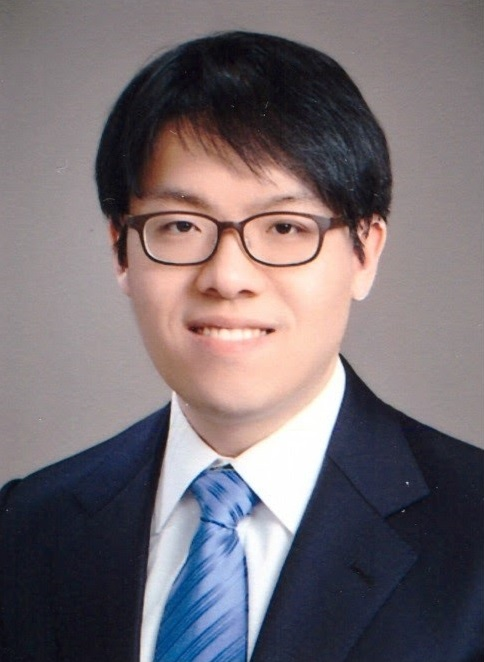}}]{SEOKJIN LEE} (Member, IEEE) received the B.S., M.S., and Ph.D. degrees in electrical and computer engineering from Seoul National University, in 2006, 2008, and 2012, respectively.
From 2012 to 2014, he was a Senior Research Engineer at LG Electronics. From 2014 to 2018, he was an Assistant Professor at the Department of Electronics Engineering, Kyonggi University, Suwon, Republic of Korea. Since 2018, he has been an Assistant Professor (promoted to an Professor, in 2025) with the School of Electrical and Electronics Engineering, Kyungpook National University, Daegu, Republic of Korea. His research interests include acoustic, sound and music signal processing, array signal processing, and blind source separation.
\end{IEEEbiography}

\vspace{11pt}

\vfill

\end{document}